# All-fiber spatiotemporally mode-locked laser with multimode fiber-based filtering

**UĞUR TEĞIN**[1,2,*]**, BABAK RAHMANI**[2]**, EIRINI KAKKAVA**[1]**, DEMETRI PSALTIS**[1] **AND CHRISTOPHE MOSER**[2]

[1]*Optics Laboratory, École Polytechnique Fédérale de Lausanne, Switzerland*
[2]*Laboratory of Applied Photonics Devices, École Polytechnique Fédérale de Lausanne, Switzerland*
*\*ugur.tegin@epfl.ch*

**Abstract:** We demonstrate the first multimode fiber interference-based filtering in an all-fiber spatiotemporally mode-locked laser. The oscillator generates dissipative soliton pulses at 1036 nm with 12 mW average power, 6.24 ps duration and 24.3 MHz repetition rate. The reported pulse energy (0.5 nJ) represents ~4 times improvement over the previously reported single-mode all-normal dispersion mode-locked lasers with multimode interference-based filtering. Numerical simulations are performed to reveal cavity and spatiotemporal mode-locking dynamics. The proposed oscillator presents an alternative approach to achieve spatiotemporal mode-locking with an all-fiber design that can be fabricated simply.

## 1. Introduction

Ytterbium-based fiber laser systems are used in optical metrology, material processing and medical applications due to its high and broadband gain [1]. Unlike the other conventionally used gain elements (Er, Tm and Ho), the emission wavelength of an Ytterbium-doped silica fiber falls in a spectral range that exhibits positive ($\beta_2 > 0$) group velocity dispersion, thus mode-locking is relatively challenging at 1 um wavelength. In the literature, by dispersion-managing with gratings or photonic crystal fibers, soliton [2], dispersion-managed soliton [3] and similariton [4] pulse types were reported with Ytterbium-based single-mode fiber lasers. Later, all-normal dispersion mode-locking with spectral filtering of chirped pulses was discovered and depending on the filtering effect, dissipative soliton [5] and amplifier similariton [6] pulses are demonstrated. Over the last decade, all-normal-dispersion fiber lasers have been studied mainly by employing dissipative soliton pulse dynamics [7]. These pulses are energy scalable and dissipative solitons with up to µJ pulse energies were demonstrated with custom made very-large mode area single-mode fibers [8].

All-fiber laser designs are a subject of high interest due to their compact and alignment-free operation. To achieve dissipative soliton pulses in all-fiber configuration various inline fiber-based filtering solutions with >6 nm bandwidth are reported in the literature [9-11]. Among them, multimode interference (MMI) based bandpass filtering is an easy and low-budget solution. Such a filter consists of a section of graded-index multimode fiber (GRIN MMF) between single or few-mode fibers and interference effects between the modes introduces frequency-dependent sinusoidal transmission which can be used as a spectral filter [12]. Recently, dissipative soliton pulse formation with 0.13 nJ energy was presented in an all-fiber configuration with the multimode interference-based bandpass filtering in a single-mode Ytterbium-based all-normal dispersion laser [13].

In the last few years, spatiotemporal mode-locking is demonstrated by harnessing the unique properties of GRIN MMF such as low modal dispersion and periodic self-imaging by Wright et al. and dissipative soliton pulses are reported [14]. With this alternative mode-locking approach in a multimode laser cavity, coherent superposition of transverse and longitudinal modes is demonstrated. By introducing spatial interactions to mode-locking mechanism complex multimode nonlinear wave propagation studies became feasible under partial feedback conditions. Later, observation of bound-state solitons and harmonic mode-locking was reported with similar cavity orientations [15,16] and all-fiber cavity with SESAM mode-locked [17]. Recently, self-similar pulse propagation is reported in spatiotemporally mode-locked multi-mode fiber laser and observed output beam quality improvements with the temporal change [18]. By tailoring spatiotemporal nonlinear pulse propagation, intracavity Kerr-induced self-beam cleaning is achieved in a multimode laser cavity with sub-100 fs pulse duration, >20 nJ pulse energy and M2 value less than 1.13 [19].

The novelty in the current paper is the use of multimode fiber interference-based filtering to construct an all-fiber spatiotemporally mode-locked laser. With this alignment-free, compact cavity design, spatiotemporally mode-locked dissipative soliton pulse generation is demonstrated. Numerical simulations are performed to reveal cavity and spatiotemporal mode-locking dynamics and led to experimental studies. The experimentally demonstrated multimode laser is self-starting and generates pulses with 0.5 nJ energy, 12 mW average power, 6.24 ps duration and 24.3 MHz repetition rate at 1036 nm. With spatiotemporal mode-locking, the achieved pulse energy represents ~4 times improvement over the previously reported single-mode mode-locked all-normal dispersion lasers with multimode interference-based filtering.

## 2. Numerical results

The schematic of the all-fiber spatiotemporally mode-locked oscillator is presented in Fig. 1a. The cavity consists of a step-index Ytterbium-based MMF segment with 10 μm core diameter, GRIN MMF segments with 50 μm core diameter and step-index passive MMF segments with 10 μm core diameter. The fiber sections with 10 μm core diameter support 3 modes and the GRIN MMF sections with 50 μm core diameter support ~240 modes at and around 1030 nm wavelength. Numerical simulations are performed to define lengths of the fiber segments, the bandwidth for multimode interference-based bandpass filtering and the possibility of spatiotemporal mode-locking in the presented cavity design.

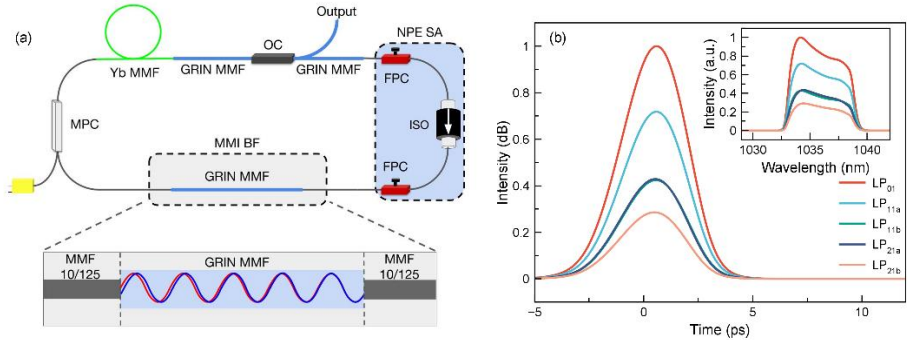

Fig. 1. (a) Schematic of the all-fiber spatiotemporally mode-locked laser with multimode fiber-based filtering: ISO, isolator; FPC, fiber polarization controller; OC, output coupler; MMI BF, multimode interference-based bandpass filter; MPC, multipump combiner. (b) Simulated mode-resolved temporal profile. Inset: Simulated mode-resolved spectral profile.

Simulations are conducted with the numerical model used by Teğin et al. [18,19]. The GRIN MMF segments are modeled with linearly polarized modes and the nonlinear multimode Schrödinger equation is simulated for these segments [20]. To decrease the computation time,

GRIN MMF segments are considered with five modes (LP01, LP11a, LP11b, LP21a and LP21b), step-index MMF segments with few-modes are considered as single-mode. Only a small portion of the modes (5 out of 250) are included in the simulation due to computational limitations however we will show that this simplified model captures essential features of the behavior of the pulse propagation inside the laser. The splice points were modeled by coupling coefficients of the modes before and after the GRIN MMF segments. The initial field in simulations is defined as a quantum noise and stable mode-locking regimes are found to be not critically dependent on the details of the coupling coefficients. The gain is modeled as Lorentzian shape with 30 dB small-signal gain and 40 nm gain bandwidth. The saturable absorber is modeled by a sinusoidal transfer function with 1 kW saturation power and 60% modulation depth.

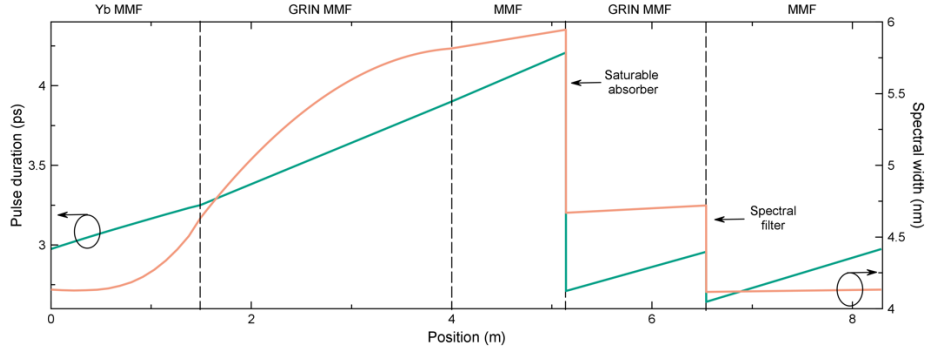

Fig. 2. Simulated pulse duration and spectral bandwidth variation over the cavity.

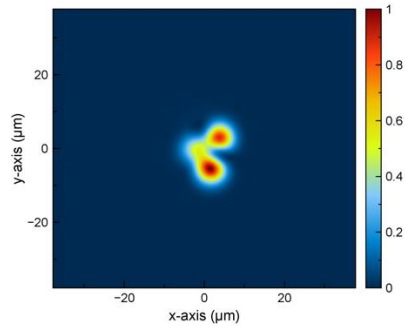

Fig. 3. Numerically obtained output beam profile.

The numerically achieved spatiotemporally mode-locked pulse shape and spectrum at the 30% output coupler are presented in Fig. 1 b. We set the excitation coefficients of the single mode section to the 5 mode GRIN MMF segments equal to [0.35, 0.25, 0.15, 0.15, 0.1] and the gain saturation energy as 1.10 nJ. It is known that in order to achieve dissipative soliton pulses in all-normal dispersion cavities bandpass filtering of chirped pulses with >6 nm bandwidth is required [5]. To ensure dissipative soliton mode-locking, we select the length of the GRIN MMF segment used for MMI filtering to be 25 cm which yields an 8 nm bandwidth bandpass filter [12]. In our simulations, dissipative soliton pulses with 0.46 nJ pulse energies and 5 nm spectral bandwidths were achieved. The output pulse duration was 3.6 ps at the output port. To understand the pulse propagation in detail, propagation of the pulse in one roundtrip is calculated and presented in Fig. 2. Due to the relatively high nonlinearity, the spectral broadening is observed in gain MMF segment which later reaches a steady-state value inside the GRIN MMF. As expected from dissipative soliton pulses in an all-normal dispersion cavity, the spectral broadening ratio in one roundtrip is small [21]. The output beam profile with the numerically calculated mode-locked field is presented in Fig. 3. Although most of the energy remains in the lower order modes, the numerically obtained output beam exhibits multimode features.

## 3. Experimental results and discussion

Encouraged by the simulations, experiments performed with the numerically designed cavity. The gain section of the oscillator is 1.5 m Yb MMF (nLight Yb-1200-10/125) pumped with a 976 nm pump diode coupled to the cavity with a pump combiner with matching passive fiber ports. The gain section is followed by a GRIN MMF based coupler with a 30% output coupling ratio. The saturable absorber is achieved as a nonlinear polarization evolution (NPE) mechanism with a polarization-sensitive inline isolator with 10 um core diameter fiber and fiber polarization controllers. Between the fibers of isolator and MPC, 25 cm GRIN MMF with 50 µm core diameter is placed to achieve MMI bandpass filtering with 8 nm bandwidth. The described oscillator is shown in Fig. 4a.

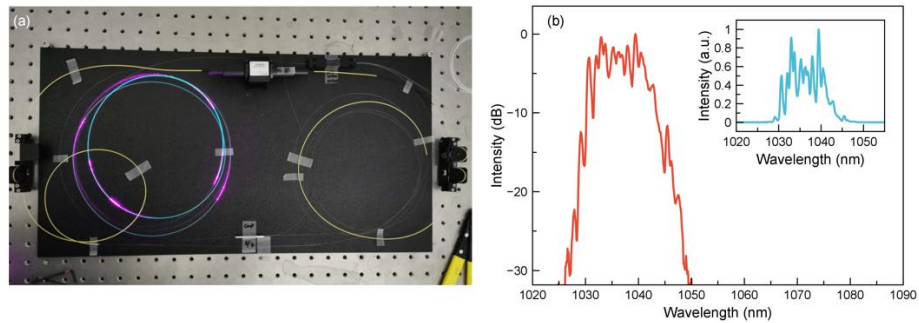

Fig. 4. (a) Experimental setup of the all-fiber spatiotemporally mode-locked laser with multimode fiber-based filtering. (b) Measured mode-locked spectrum in logarithmic scale. Inset: Measured mode-locked spectrum in linear scale.

Spatiotemporal mode-locking is achieved easily by adjusting the intracavity polarization with fiber polarization controllers. The recorded spectra from the output coupler are presented in Fig. 4b with spectral 10 nm spectral width at 1036 nm central wavelength. The presented spectrum is measured with a 0.5 nm resolution and features a jagged profile. Similar behavior is reported for spatiotemporally mode-locked lasers and lasers with MMI segments. Since the reported cavity operates based on these cases simultaneously, the reason for the jagged spectrum can be related to the aforementioned operation type and filtering. The self-starting mode-locking operation of a single-pulse train with 24.29 MHz is presented in Fig.5 a. The output power of the laser is measured as 12 mW which corresponds to ~0.5 nJ pulse energy. The temporal characterizations are performed with second-order nonlinear autocorrelation. The laser produces chirped pulses with 6.24 ps pulse duration (see Fig. 5b).

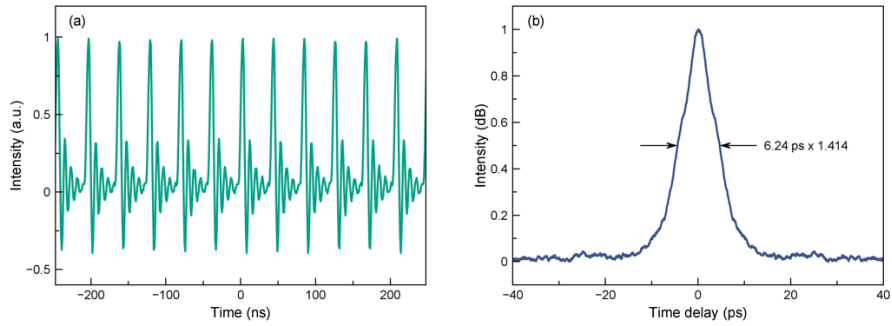

Fig. 5. (a) Measured single-pulse train of the spatiotemporally mode-locked laser. (b) Autocorrelation trace of the chirped pulse obtained from 50 cm GRIN MMF of output coupler.

It is important to consider the effect of the fiber length and diameter of the output coupler to laser output properties. The pulses propagate through 50 cm GRIN MMF with 50 um core diameter after leaving the oscillator. This output fiber causes highly multimode propagation to 0.5 nJ pulses and as presented in Fig. 6a, the near field measurement of the output beam profile is speckled. In addition to its spatial effect, such a multimode propagation can cause temporal changes as well. Numerical simulations suggest an output pulse duration ~4 ps but experimental measurements indicated around 6 ps larger pulse duration. This 2 ps difference can be the result of the highly multimode propagation caused by the output coupler in addition to the differences caused by the simplified numerical model. The oscillator is also characterized in the frequency domain with a radio frequency for stability purposes. The fundamental repetition rate of the laser is verified with a radio frequency (RF) analyzer (HP 3585A) as 24.29 MHz. With 1 kHz span and 10 Hz resolution bandwidth, a sideband suppression ratio around 70 dB is measured (Fig. 6b). The fiber laser has outstanding stability both in the short and long term. The laser continues the mode-locking operation uninterrupted for weeks, without a sign of degradation.

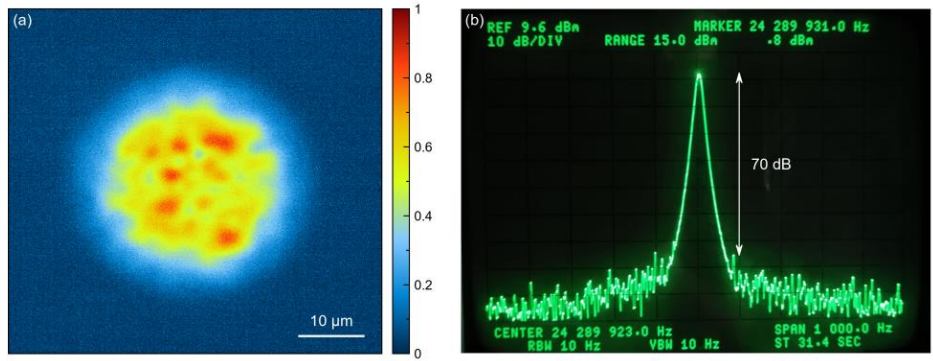

Fig. 6. (a) Measured near-field beam profile from 50 cm GRIN MMF of output coupler. (b) Measured radio frequency spectrum with 1 kHz span and 10 Hz resolution bandwidth.

## 4. Conclusion

In conclusion, we numerically and experimentally demonstrate an all-fiber spatiotemporally mode-locked laser with multimode fiber interference-based filtering. The Ytterbium-based all-normal dispersion multimode oscillator generates 6.24 ps pulses with 0.5 nJ pulse energy, 12 mW average power and 24.3 MHz repetition rate. Compared to the Ytterbium-based single-mode mode-locked lasers with multimode interference-based filtering, the reported spatiotemporally mode-locked laser produces ~4 times more powerful pulses. The all-fiber cavity design provides high stability due to the inherent alignment-free construction. We believe the proposed cavity presents an alternative approach to achieve spatiotemporal mode-locking with a simple, all-fiber design that can be used when a clean Gaussian beam is not required such as speckle interferometry and structured illumination applications.